# Second quantization and gauge invariance.

Dan Solomon
Rauland-Borg Corporation
Mount Prospect, IL
Email: dsolom2@uic.edu

June 30, 2013.

## Abstract.

It is well known that the single particle Dirac equation is gauge invariant. This means that observable quantities, such as the current density, are not affected by a gauge transformation. However what happens when the method of "second quantization" is applied to convert a single particle theory into a field theory? In this case it will be shown that the theory is no longer gauge invariant. This will be shown by considering the second quantization of a zero mass Dirac field in 1+1 dimensions and examining the change of the current density operator due to a gauge transformation.

## 1. Introduction.

Quantum field theory is generally assumed to be gauge invariant [1,2]. A change in the gauge is a change in the electromagnetic potential that doesn't produce a change in the electric or magnetic field. Such a change should not produce any change in a physically observably quantity such as the current or charge density. However it is well known that when the vacuum current is calculated using standard perturbation theory the results are not gauge invariant. Non-gauge invariant terms appear in the results which must be removed to yield a physically correct result.

For an example of this consider Section 14.2 of Greiner et al [2] where the solution for the vacuum polarization tensor is given by,

$$\pi^{\mu\nu}(k) = \left(g^{\mu\nu}k^2 - k^\mu k^\nu\right)\pi(k^2) + g^{\mu\nu}\pi_{sp}(k^2) \qquad (1.1)$$

where $\pi(k^2)$ and $\pi_{sp}(k^2)$ are defined in Ref. [2]. As discussed in [2] the first term to the right of the equals sign is gauge invariant however the second term is not unless $\pi_{sp}(k^2)$ is zero. It is shown by Greiner et al that this is not the case. Therefore this second term must be removed in order to get a physically correct result.

Another instance of this type of problem is provided in Section 6.4 of Nishijima [3] where an expression for the vacuum polarization tensor is derived and is shown to include non-gauge invariant terms which must be dropped form the expression to obtain the "correct" gauge invariant result. A number of additional examples and a detailed discussion of this problem is given in Ref [4,5].

The purpose of this paper is to show that the failure of gauge invariance occurs during the process of second quantization of the quantum field. We will do this by examining the effect of a gauge transformation on the current operator of a second quantized Dirac field in 1+1 dimensions with zero mass fermions.

## 2. Gauge invariance of the single particle current density.

The Dirac equation for a single massless fermion in 1+1 dimensions in the presence of an external electric potential, $(A_0(z,t), A_1(z,t))$, is,

$$i\frac{\partial \varphi(z,t)}{\partial t} = H\varphi(z,t) \tag{2.1}$$

where,

$$H = H_0 - \sigma_3 A_1(z,t) + A_0(z,t) \tag{2.2}$$

and,

$$H_0 = -i\sigma_3 \frac{\partial}{\partial z} \tag{2.3}$$

where $\sigma_3$ is the Pauli matrix with $\sigma_3 = \begin{pmatrix} 1 & 0 \\ 0 & -1 \end{pmatrix}$.

The electric field is given by,

$$E = -\left(\frac{\partial A_1}{\partial t} + \frac{\partial A_0}{\partial z}\right) \tag{2.4}$$

A gauge transformation is a change in the electric potential that does not produce a change in the electric field. Such a change is given by,

$$(A_0, A_1) \to (A_0 - \partial \chi/\partial t, A_1 + \partial \chi/\partial z). \tag{2.5}$$

where $\chi(z,t)$ is an arbitrary function.

The solution of (2.1) can be easily shown to be,

$$\varphi(z,t) = W(z,t)\varphi_0(z,t) \tag{2.6}$$

where $\varphi_0(z,t)$ is the solution to the free field Dirac equation,

$$i\frac{\partial \varphi_0(z,t)}{\partial t} = H_0 \varphi_0(z,t) \tag{2.7}$$

and can be written as,

$$\varphi_0(z,t) = e^{-iH_0 t}\varphi_0(z). \tag{2.8}$$

The quantity $W(z,t)$ is given by,

$$W(z,t) = \begin{pmatrix} e^{-ic_1} & 0 \\ 0 & e^{-ic_2} \end{pmatrix} \tag{2.9}$$

where $c_1(z,t)$ and $c_2(z,t)$ satisfy the following differential equations,

$$\frac{\partial c_1}{\partial t} + \frac{\partial c_1}{\partial z} = A_0 - A_1 \tag{2.10}$$

and,

$$\frac{\partial c_2}{\partial t} - \frac{\partial c_2}{\partial z} = A_0 + A_1. \tag{2.11}$$

Let,

$$A_0 = -\frac{\partial \chi(z,t)}{\partial t} \text{ and } A_1 = \frac{\partial \chi(z,t)}{\partial z} \tag{2.12}$$

Use this in (2.4) to obtain $E = 0$. Therefore (2.12) is a gauge transformation from zero electric field. Use (2.12) in (2.10) and (2.11) to obtain,

$$c_1(z,t) = c_2(z,t) = -\chi(z,t). \tag{2.13}$$

Use this in (2.6) along with (2.9) and (2.8) to obtain,

$$\varphi(z,t) = e^{+i\chi(z,t)} e^{-iH_0 t}\varphi_0(z). \tag{2.14}$$

The current density for a single fermion is defined by,

$$J(z,t) = \varphi^\dagger(z,t)\sigma_3 \varphi(z,t). \tag{2.15}$$

To determine the impact of a gauge transformation on the current density use (2.14) along with $\varphi^\dagger(z,t) = \left(e^{-iH_0 t}\varphi_0(z)\right)^\dagger e^{-i\chi(z,t)}$ in the above expression for $J(z,t)$ to obtain,

$$J(z,t) = \left(e^{-iH_0 t}\varphi_0(z)\right)^\dagger \sigma_3 \left(e^{-iH_0 t}\varphi_0(z)\right). \tag{2.16}$$

We see that the dependence on the function $\chi(z,t)$ does not appear in the above expression. Therefore the gauge transformation does not change the current density which proves that the current density for a single fermion is gauge invariant. This is, of course, a standard result.

## 3. Second quantization.

As we have just shown the current density for a single fermion is gauge invariant. The next step is to apply the usual methods of second quantization and to determine whether or not the resulting quantum field theory is gauge invariant. We will follow the approach of Ref [6] in which a second quantized formulation for massless fermions interacting with an external field in a two dimensional model was discussed.

Let $\phi_{\lambda,p}(z)$ be the eigenfunctions of the free field Hamiltonian with energy eigenvalues $\varepsilon_{\lambda,p}$. They satisfy the relationship,

$$H_0 \phi_{\lambda,p}(z) = \varepsilon_{\lambda,p} \phi_{\lambda,p}(z) \tag{3.1}$$

where,

$$\phi_{\lambda,p}(z) = \frac{1}{2\sqrt{L}} \begin{pmatrix} 1 + \dfrac{\lambda p}{|p|} \\ 1 - \dfrac{\lambda p}{|p|} \end{pmatrix} e^{ipz}; \quad \varepsilon_{\lambda,p} = \lambda |p| \tag{3.2}$$

and where $\lambda = \pm$ is the sign of the energy, $p$ is the momentum, and $L$ is the 1 dimensional integration volume. We assume periodic boundary conditions so that the momentum $p = 2\pi r/L$ where $r$ is an integer. According to the above definitions the quantities $\phi_{-,p}(z)$ are negative energy states with energy $\varepsilon_{-,p} = -|p|$ and the quantities $\phi_{+,p}(z)$ are positive energy states with energy $\varepsilon_{+,p} = |p|$.

The $\varphi_{\lambda,p}^{(0)}(z)$ form an orthonormal basis set and satisfy,

$$\int \phi^\dagger_{\lambda,p}(z)\phi_{\lambda',p'}(z)dz = \delta_{\lambda\lambda'}\delta_{pp'} \tag{3.3}$$

where integration from $-L/2$ to $+L/2$ is implied.

Define the project operators $P^0_\pm$ where $P^0_-$ projects into the negative energy states and $P^0_+$ projects into the positive energy states. The projection operators are defined by their action on a function $f(x)$:

$$P^0_\pm f = \sum_k \phi_{\pm,k}\langle\phi_{\pm,k},f\rangle. \tag{3.4}$$

Following Ref. [6] define the field operator,

$$\psi(f) = b(P^0_+ f) + d^\dagger(\overline{P^0_- f}). \tag{3.5}$$

The operators $b(P^0_+ f)$ and $d(P^0_- f)$ satisfy the canonical anticommutation relationships (CAR),

$$\{b(P^0_+ f), b^\dagger(P^0_+ g)\} = \langle f, P^0_+ g\rangle, \quad \{d(P^0_- f), d^\dagger(P^0_- g)\} = \langle f, P^0_- g\rangle. \tag{3.6}$$

with all other CARs equal to zero. These operators act on a Fock space $\Im(H)$. The vacuum state $\Omega_0 \in \Im(H)$ is annihilated by $b$ and $d$:

$$b(P^0_- f)\Omega_0 = 0, \quad d(P^0_- f)\Omega_0 = 0. \tag{3.7}$$

Consider a unitary operator $V$ that acts on the single particle wave function $\varphi(x)$ such that $\varphi_V(x) = V\varphi(x)$. How does this unitary operator impact the Fock space? If this $V$ satisfies a certain condition then the Fock space will be acted by the operator $\Gamma(V)$ where $\Gamma(V)$ acts on the field operator according to:

$$\psi(Vf) = \Gamma(V)\psi(f)\Gamma^\dagger(V). \tag{3.8}$$

It is has been shown that in order for $\Gamma(V)$ to exist the unitary operator $V$ must be Hilbert-Schmidt [7]. In this case the operator $V$ is said to be unitary implementable.

## 4. Failure of gauge invariance.

In Ref [6] a generalized charge operator $Q(A)$ is defined,

$$Q(A) = \sum_{n,m}\left(b^\dagger_n A^{++}_{nm} b_m + b^\dagger_n A^{+-}_{nm} d^\dagger_m + d_n A^{-+}_{nm} b_m - d^\dagger_m A^{--}_{nm} d_n\right) \tag{4.1}$$

where $A$ is a bounded operator on the Hilbert space and where,

$$A_{nm}^{\pm\pm} = \langle \phi_{\pm,n}, A\phi_{\pm,m} \rangle , \quad A_{nm}^{\pm\mp} = \langle \phi_{\pm,n}, A\phi_{\mp,m} \rangle, \quad b_n = b(\phi_{+,n}), \quad d_n = d(\phi_{-,n}) \quad (4.2)$$

If $V$ is a unitary operator acting on the Hilbert space and $\Gamma(V)$ is the associated second quantized operator then it is shown in [6] that,

$$\Gamma(V)Q(A)\Gamma^\dagger(V) = Q(VAV^\dagger) + \Delta(A) \quad (4.3)$$

where,

$$\Delta(A) = Tr(P_+^0 A P_- P_+^0) - Tr(P_-^0 A P_+ P_-^0) \quad (4.4)$$

with,

$$P_\pm = V^\dagger P_\pm^0 V . \quad (4.5)$$

A formal expression for the current density operator smeared over a real-valued function $f(x)$ is given in Ref [8] by,

$$J(f) = \int :\psi^\dagger(z)\sigma_3\psi(z): f(z) dz . \quad (4.6)$$

Using this relationship as a model and referring to (4.1) we find that the second quantized current density operator smeared over $f(z)$ is given by $Q(\sigma_3 f)$.

The effect of the gauge transformation is to act on the Hilbert space with the unitary operator $V_\chi = e^{i\chi}$. It has been shown that this operator is unitary implementable [8, 9]. Use this in (4.3) to show that the effect of the gauge transformation on the current operator is,

$$Q_\chi(\sigma_3 f) = \Gamma(e^{i\chi})Q(\sigma_3 f)\Gamma^\dagger(e^{i\chi}) = Q(e^{i\chi}\sigma_3 f e^{-i\chi}) + \Delta_\chi(\sigma_3 f) . \quad (4.7)$$

where,

$$\Delta_\chi(\sigma_3 f) = Tr(P_+^0 \sigma_3 f e^{-i\chi} P_-^0 e^{+i\chi} P_+^0) - Tr(P_-^0 \sigma_3 f e^{-i\chi} P_+^0 e^{+i\chi} P_-^0) . \quad (4.8)$$

The gauge transformation has taken current operator $Q(\sigma_3 f)$ into the current density operator $Q_\chi(\sigma_3 f)$. If the current density operator is gauge invariant then the quantity $\chi$ must disappear from the right hand side of (4.7).

Consider the term $Q(e^{i\chi}\sigma_3 f e^{-i\chi})$. Since $e^{i\chi}\sigma_3 f e^{-i\chi} = \sigma_3 f$ it is evident that $Q(e^{i\chi}\sigma_3 f e^{-i\chi}) = Q(\sigma_3 f)$. Therefore this part of the expression is independent of $\chi$. Next we have to evaluate $\Delta_\chi(\sigma_3 f)$. In the Appendix it is shown that,

$$\Delta_\chi(\sigma_3 f) = -\frac{1}{\pi}\int dz f(z)\frac{d\chi(z)}{dz}. \tag{4.9}$$

Use this in (4.7) to obtain,

$$Q_\chi(\sigma_3 f) = Q(\sigma_3 f) - \frac{1}{\pi}\int dz f(z)\frac{d\chi(z)}{dz}. \tag{4.10}$$

The last term in the above expression is, in general, non-zero and is dependent on $\chi(z)$. Therefore the current density operator is not gauge invariant.

## 5. Conclusion.

We have examined the effect of a gauge transformation on the current density for zero mass Dirac field in 1+1 dimensions. This problem was motivated by the fact that standard calculations of the vacuum current density in quantum field theory yield non-gauge invariant results. It was shown that for a single fermion the current density is, indeed, gauge invariant. However when the method second quantization is used to produce a field theory then the result is no longer gauge invariant. This, then, explains why calculations of the vacuum current using perturbation theory do not produce gauge invariant results and must be "corrected" by removing the non-gauge invariant terms.

## Appendix.

In the following we will evaluate $\Delta_\chi(\sigma_3 f)$. Use (3.4) and (4.5) to obtain,

$$Tr(P_+^0 A P_- P_+^0) = \sum_p \sum_k \langle\phi_{+,p}, AV^\dagger\phi_{-,k}\rangle\langle\phi_{-,k}, V\phi_{+,p}\rangle. \tag{5.1}$$

This can be rewritten as,

$$Tr(P_+^0 A P_- P_+^0) = TR\iint A(z)V^\dagger(z)\left(\sum_k \phi_{-,k}(z)\phi_{-,k}^\dagger(z')\right)V(z')\left(\sum_p \phi_{+,p}(z')\phi_{+,p}^\dagger(z)\right)dzdz'. \tag{5.2}$$

where $TR$ is used to symbolize the trace operation over the spinnor indices only. Use (3.2) to obtain,

$$\sum_{k} \phi_{-,k}(z) \phi_{-,k}^{\dagger}(z') = \frac{1}{L} \sum_{k \geq 0} \left[ \begin{pmatrix} 0 & 0 \\ 0 & 1 \end{pmatrix} e^{ik(z-z')} + \begin{pmatrix} 1 & 0 \\ 0 & 0 \end{pmatrix} e^{-ik(z-z')} \right] \quad (5.3)$$

and,

$$\sum_{p} \phi_{+,p}(z) \phi_{+,p}^{\dagger}(z') = \frac{1}{L} \sum_{p \geq 0} \left[ \begin{pmatrix} 1 & 0 \\ 0 & 0 \end{pmatrix} e^{ip(z'-z)} + \begin{pmatrix} 0 & 0 \\ 0 & 1 \end{pmatrix} e^{-ip(z'-z)} \right]. \quad (5.4)$$

Next use $A = \sigma_3 f$ and $V(z) = e^{i\chi(z)}$ along with the above relationships to obtain,

$$Tr\left(P_+^0 A P_- P_+^0\right) = \frac{1}{L^2} \iint dz' dz f(z) e^{-i\chi(z)} e^{i\chi(z')} B(z-z') \quad (5.5)$$

where,

$$B(z-z') = \sum_{k,p>0} \left( e^{-i(p+k)(z-z')} - e^{i(p+k)(z-z')} \right). \quad (5.6)$$

Use $p = 2\pi n/L$ and $k = 2\pi m/L$ to obtain,

$$B(z-z') = \sum_{n=0}^{\infty} \sum_{m=0}^{\infty} \left( \exp\left[-i\frac{2\pi}{L}(n+m)(z-z')\right] - \exp\left[i\frac{2\pi}{L}(n+m)(z-z')\right] \right). \quad (5.7)$$

Now take $L \to \infty$ and use,

$$\sum_{n=0}^{\infty} g\left(\frac{2\pi n}{L}\right) \to \frac{L}{2\pi} \int_0^{\infty} g(\omega) d\omega \quad (5.8)$$

in (5.7) to obtain,

$$B(z-z') = \left(\frac{L}{2\pi}\right)^2 \int_0^{\infty} d\omega' \int_0^{\infty} d\omega \left( \exp\left[-i(\omega'+\omega)(z-z')\right] - \exp\left[i(\omega'+\omega)(z-z')\right] \right) \quad (5.9)$$

To evaluate this use $s = \omega' + \omega$ and $s' = \omega' - \omega$ in the above to obtain

$$B(z-z') = \frac{1}{2}\left(\frac{L}{2\pi}\right)^2 \int_0^{\infty} ds \int_{-s}^{+s} ds' \left( \exp\left[-is(z-z')\right] - \exp\left[is(z-z')\right] \right). \quad (5.10)$$

This yields,

$$B(z-z') = \left(\frac{L}{2\pi}\right)^2 \int_0^{\infty} s ds \left( \exp\left[-is(z-z')\right] - \exp\left[is(z-z')\right] \right). \quad (5.11)$$

This can be further evaluated to obtain,

$$B(z-z') = i\left(\frac{L}{2\pi}\right)^2 \frac{d}{dz} \int_{-\infty}^{\infty} \exp\left[-is(z-z')\right] ds. \quad (5.12)$$

Use,

$$\int_{-\infty}^{\infty} \exp\left[-is(z-z')\right]ds = 2\pi\delta(z-z') \tag{5.13}$$

in (5.12) to obtain,

$$B(z-z') = \frac{iL^2}{2\pi}\frac{d}{dz}\delta(z-z'). \tag{5.14}$$

Therefore,

$$Tr\left(P_+^0 A P_- P_+^0\right) = \frac{i}{2\pi}\iint dz'dz f(z) e^{-i\chi(z)} e^{i\chi(z')} \frac{d}{dz}\delta(z-z'). \tag{5.15}$$

This is evaluated to obtain,

$$Tr\left(P_+^0 A P_- P_+^0\right) = -\frac{1}{2\pi}\int dz f(z)\frac{d\chi(z)}{dz}. \tag{5.16}$$

Similarly it can be shown that,

$$Tr\left(P_-^0 A P_+ P_-^0\right) = -Tr\left(P_+^0 A P_- P_+^0\right). \tag{5.17}$$

Use (5.16) and (5.17) along with (4.4) to obtain,

$$\Delta_\chi(\sigma_3 f) = -\frac{1}{\pi}\int dz f(z)\frac{d\chi(z)}{dz}$$

which is Eq. (4.9) in the text.